\documentclass[aps,prd,groupedaddress,amssymb,twocolumn,eqsecnum,showpacs,epsfig,nofootinbib]{revtex4}

\usepackage{graphicx}
\usepackage{bm}
\usepackage{dcolumn}
\usepackage{amsmath}

\usepackage{amsmath}
\usepackage{amssymb}

\numberwithin{equation}{section}
\usepackage{epsfig}

\begin{document}

\newcommand{\newc}{\newcommand}

\newc{\be}{\begin{equation}}
\newc{\ee}{\end{equation}}
\newc{\ba}{\begin{eqnarray}}
\newc{\ea}{\end{eqnarray}}
\newc{\bea}{\begin{eqnarray*}}
\newc{\eea}{\end{eqnarray*}}
\newc{\D}{\partial}
\newc{\ie}{{\it i.e.} }
\newc{\eg}{{\it e.g.} }
\newc{\etc}{{\it etc.} }
\newc{\etal}{{\it et al.}}
\newc{\lcdm}{$\Lambda$CDM}
\newcommand{\nn}{\nonumber}
\newc{\ra}{\rightarrow}
\newc{\lra}{\leftrightarrow}
\newc{\lsim}{\buildrel{<}\over{\sim}}
\newc{\gsim}{\buildrel{>}\over{\sim}}

\title{The spherical collapse model and cluster formation beyond the $\Lambda$ cosmology:
Indications for a clustered dark energy?}

\author{Spyros Basilakos}
\affiliation{Academy of Athens, Research Center for Astronomy and Applied Mathematics,
 Soranou Efesiou 4, 11527, Athens, Greece}
 \author{Juan Carlos Bueno Sanchez and Leandros Perivolaropoulos,}
\affiliation{Department of Physics, University of Ioannina, Greece}

\begin{abstract}
We generalize the small scale dynamics of the universe by
taking into account models with an
equation of state which evolves with time, and provide a complete
formulation of the cluster virialization attempting to address the
nonlinear regime of structure formation.
In the context of the current dark energy models,
we find that galaxy clusters appear to form at $z\sim 1-2$, in agreement
with previous studies. Also, we investigate thoroughly the evolution of
spherical matter perturbations, as the latter decouple from the background
expansion and start to ``turn around'' and finally collapse.
Within this framework, we find that the concentration parameter depends on the choice
of the considered dark energy (homogeneous or clustered). In particular,
if the distribution of the dark energy is clustered then
we produce more concentrated structures with respect to the homogeneous dark energy.
Finally, comparing the predicted concentration parameter with the observed
concentration parameter, measured for four massive galaxy clusters, we find that
the scenario which contains a pure homogeneous dark energy is unable to reproduce
the data. The situation becomes somewhat better in the case of an
inhomogeneous (clustered) dark energy.

 \end{abstract}
\maketitle

\section{Introduction}
Recent studies in observational cosmology lead to the conclusion that
the available high quality cosmological data
(Type Ia supernovae, CMB, etc.) are
well fitted by an emerging ``standard model''. This standard
model, assuming flatness, is described by the Friedman equation
\be
H^2(a)=\left(\frac{{\dot a}}{a}\right)^2=\frac{8 \pi G}{3}\left[\rho_{m}(a)+
\rho_{X}(a)\right]
\label{fe1}
\ee
where $a(t)$ is the
scale factor of the universe, $\rho_{m}(a)$ is the density
corresponding to the sum of baryonic and cold dark matter,
with the latter needed to explain clustering, and an extra component
$\rho_{X}(a)$ with negative pressure called dark energy needed to
explain the observed accelerated cosmic expansion
(\cite{Riess07,Spergel07,essence,komatsu08,Kowal08} and references therein).

The nature of the dark energy is one of the most fundamental and
difficult problems in physics and cosmology. Indeed, during the
last decade there has been a theoretical debate among the cosmologists
regarding the nature of the
exotic ``dark energy''. Various candidates have been proposed in the
literature, such as a cosmological constant $\Lambda$ (vacuum),
time-varying $\Lambda(t)$ cosmologies, quintessence, $k-$essence,
vector fields, phantom, tachyons, Chaplygin gas and the list goes on
(see \cite{Ratra88,Oze87,Weinberg89,Wetterich:1994bg,Caldwell98,Brax:1999gp,
KAM,fein02,Caldwell,Peebles03,chime04,Brookfield:2005td,Copel06,Boehmer:2007qa,Friem08}
and references therein).
Within this framework, high energy field
theories generically indicate that the equation of state of such a
dark energy is a function of the cosmic time. To identify this type
of evolution of the equation of state, a detailed form of the
observed $H(a)$ is required which may be obtained by a combination
of multiple dark energy probes.

A serious issue here is how (and when) the large scale
structures and in particular galaxy clusters form. Galaxy
clusters occupy an eminent position in the structure
hierarchy, being the most massive virialized systems known
and therefore they appear to be ideal tools for testing
theories of structure formation and extracting cosmological
information.
The cluster distribution basically traces scales that have
not yet undergone the non-linear phase of gravitationally
clustering, thus simplifying their connections
to the initial conditions of cosmic structure formation.
In the last decade many authors have been involved in this kind of studies
and have found that the main features
of the large scale structures (formation epoch, shape etc)
can potentially affected by the dark energy
\cite{Lahav91,Wang98,Iliev01,Battye03,Bas03,manera,Wein03,Mota04,
Horel05,Zeng05,Maor05,david,Perc05,Nunes06,Wang06,Basi07}.

The aim of this work is along the same lines, attempting to
investigate the main properties of the non-linear spherical model
for a large family of dark energy models, in which the
corresponding equation of state parameter varies as a function of
the scale factor of the universe, $w=w(a)$. The plan of the paper
is as follows. In section 2 we briefly discuss the dark energy
issue. In section 3 we present the various dark energy models and
in section 4 we use a joint statistical analysis, in order to
place constraints on the main cosmological parameters. Section 5
outlines the theoretical analysis of the spherical collapse model
in which the equation of state parameter varies with the cosmic
time. In section 6 we present the corresponding theoretical
predictions regarding the formation of the galaxy clusters. In
section 7, we compare the predicted concentration parameters with
those found by four galaxy clusters at relatively large redshifts
$0.18\leq z \le 0.45$), using the Subaru 8.3 telescope
\cite{Miyazaki:2002wa}. Finally, we draw our conclusions in
section 8. Throughout the paper we will use
$H_{0}=70.5$km/sec/Mpc.

\section{The dark energy equation of state}
In the framework of the general relativity it is well known that
for homogeneous and isotropic flat cosmologies, driven by
non-relativistic matter and dark energy with equation
of state $p_X=w(a)\rho_X$, the expansion rate of the Universe
can be written as (see eq.\ref{fe1})
\begin{equation}
E^{2}(a)=\frac{H^{2}(a)}{H_{0}^{2}}=
\Omega_{m}a^{-3}+\Omega_{X}{\rm e}^{3\int^{1}_{a} d{\rm lny}[1+w(y)]} \;\;.
\label{nfe1}
\end{equation}
Note, that $E(a)$ is the normalized Hubble flow,
$\Omega_{m}$ is the dimensionless matter density
at the present epoch, $\Omega_{X}=1-\Omega_{m}$ is the
corresponding dark energy density parameter
and $w(a)$ is the dark energy equation
of state. Inverting the above equation we simply derive
\begin{equation}
\label{eos22}
w(a)=\frac{-1-\frac{2}{3}a\frac{{d\rm lnE}}{da}}
{1-\Omega_{m}a^{-3}E^{-2}(a)} \;\;.
\end{equation}
The exact nature of the dark energy has yet to be found and thus
the dark energy equation of state
parameter includes our ignorance regarding the
physical mechanism which leads to a late cosmic acceleration.

On the other hand, it is possible to extent the previous
methodology in the framework of modified gravity (see
\cite{Linjen03, Linder2004}). Instead of using the exact Hubble
flow through a modification of the Friedmann equation we can
utilize a Hubble flow that looks like the nominal one (see
eq.\ref{fe1}). The key point here is to consider that the
accelerated expansion of the universe can be attributed to a
``geometrical'' dark energy component. Due to the fact that the
matter density in the universe (baryonic+dark) can not accelerate
the cosmic expansion we perform the following parametrization
\cite{Linjen03, Linder2004}:
\begin{equation}
E^{2}(a)=\frac{H^{2}(a)}{H_{0}^{2}}=
\Omega_{m}a^{-3}+\delta H^{2} \;\;.
\label{nfe2}
\end{equation}
Obviously, with the aid of the latter approach we include any
modification to the Friedmann equation of general relativity in
the last term of eq.(\ref{nfe2}). Now from eqs.(\ref{eos22},
\ref{nfe2}) we can derive, after some algebra, the ``geometrical''
dark energy equation of state
\begin{equation}
\label{eos222}
w(a)=-1-\frac{1}{3}\;\frac{d{\rm ln}\delta H^{2}}{d{\rm ln}a} \;\;.
\end{equation}
From now on, for the modified
cosmological models we will use the above formulation.

\section{Likelihood Analysis}
In this work we use various cosmological observations in order to
constrain the dark energy models, explored here (see section 3).
First of all, we use the  Baryonic Acoustic Oscillations
(BAOs). BAOs are produced by pressure (acoustic) waves in the
photon-baryon plasma in the early universe, generated by dark
matter overdensities. Evidence of this excess was recently found
in the clustering properties of the luminous SDSS red-galaxies
\cite{Eis05} and it can provide a ``standard ruler'' with which we
can constraint the dark energy models. In particular, we use the
following estimator:
\begin{equation}
A({\bf p})=\frac{\sqrt{\Omega_{m}}}{[z^{2}_{s}E(a_{s})]^{1/3}}
\left[\int_{a_{s}}^{1} \frac{da}{a^{2}E(a)}
\right]^{2/3}
\end{equation}
measured from the SDSS data to be $A=0.469\pm 0.017$, where $z_{s}=0.35$
[or $a_{s}=(1+z_{s})^{-1}\simeq 0.75$] and $E(a)\equiv H(a)/H_0$ is the
normalized Hubble flow.
Therefore, the corresponding $\chi^{2}_{\rm BAO}$ function is simply written
\begin{equation}
\chi^{2}_{\rm BAO}({\bf p})=\frac{[A({\bf p})-0.469]^{2}}{0.017^{2}}
\end{equation}
where ${\bf p}$ is a vector containing the cosmological
parameters that we want to fit.

On the other hand, a very accurate and deep geometrical probe of dark energy is the
angular scale of the sound horizon at the last scattering surface as
encoded in the location $l_1^{TT}$ of the first peak of the
Cosmic Microwave Background (CMB)
temperature perturbation spectrum. This probe is described by the  CMB shift parameter
\cite{Bond:1997wr,Trotta:2004qj,Nesseris:2006er} which is defined as
\be
\label{shift}
R=\sqrt{\Omega_{m}}\int_{a_{ls}}^1 \frac{da}{a^2
E(a)} \;\;.
\ee
The shift parameter measured from the WMAP 5-years
data \cite{komatsu08} to be $R=1.71\pm 0.019$ at $z_{ls}=1090$
[or $a_{ls}=(1+z_{ls})^{-1}\simeq 9.17\times 10^{-4}$].
In this case, the $\chi^{2}_{\rm cmb}$ function is given
\begin{equation}
\chi^{2}_{\rm cmb}({\bf p})=\frac{[R({\bf p})-1.71]^{2}}{0.019^{2}}
\end{equation}

Finally, we utilize the Union08 sample of
307 supernovae of Kowalski et al. \cite{Kowal08}.
In this case, the $\chi^{2}_{\rm SNIa}$ function becomes:
\begin{equation}
\label{chi22}
\chi^{2}_{\rm SNIa}({\bf p})=\sum_{i=1}^{307} \left[ \frac{ {\cal \mu}^{\rm th}
(a_{i},{\bf p})-{\cal \mu}^{\rm obs}(a_{i}) }
{\sigma_{i}} \right]^{2} \;\;.
\end{equation}
where $a_{i}=(1+z_{i})^{-1}$ is the observed scale factor of
the Universe, $z_{i}$ is the observed redshift, ${\cal \mu}$ is the
distance modulus ${\cal \mu}=m-M=5{\rm log}d_{L}+25$
and $d_{L}(a,{\bf p})$ is the luminosity distance
\begin{equation}
d_{L}(a,{\bf p})=\frac{c}{a} \int_{a}^{1} \frac{{\rm d}y}{y^{2}H(y)}
\end{equation}
where $c$ is the speed of light.
We can combine the above probes by using a joint likelihood analysis:
$${\cal L}_{tot}({\bf p})=
{\cal L}_{\rm BAO} \times {\cal L}_{\rm cmb}\times {\cal L}_{\rm SNIa} $$
or
$$\chi^{2}_{tot}({\bf p})=\chi^{2}_{\rm BAO}+\chi^{2}_{\rm cmb}+\chi^{2}_{\rm SNIa}$$
in order to put even further constraints on the parameter space used.
Note, that we define the likelihood estimator \footnote{Likelihoods
are normalized to their maximum values. Note, that the step of sampling is 0.01
and the errors of the fitted
parameters represent $2\sigma$ uncertainties.
Note that the overall number of
data points used is $N_{tot}=309$ and the degrees of freedom:
$dof= N_{tot}-n_{\rm fit}$, with $n_{\rm fit}$ the number of fitted
parameters, which vary for the different models.}
as: ${\cal L}_{j}\propto {\rm exp}[-\chi^{2}_{j}/2]$.

\section{Constraints on the flat dark energy models}\label{sec.constr}
In this section, we consider a large family of flat dark energy
models and with the aid of the above observational data we attempt
to put tight constraints on their free parameters. In the
following subsections, we briefly present these cosmological
models which trace differently the nature of the dark energy.

\subsection{Constant equation of state - QP model}
In this case the equation of state is constant (see for a review,
\cite{Peebles03}; hereafter QP-models).
Such dark energy models do not have much physical motivation. In
particular, a constant equation of state parameter requires a fine
tunning of the potential and kinetic energies of the real scalar
field. Despite the latter problem, these dark energy models
have been used in the literature due to their simplicity.
Notice, that dark energy models with a canonical kinetic term
have $-1\le w<-1/3$. On the other hand, models of phantom dark energy
($w<-1$) require exotic nature, such as a scalar field with a negative
kinetic energy.
%If $-1\le w<-1/3$ we have the quintessence models.
%Such cosmological models are based on the general assumption that
%the real scalar field $\phi$ rolls down the potential $V(\phi)$
%and therefore it could resemble the dark energy.
%On the other hand, if $w<-1$ then we get the  Phantom
%cosmologies \cite{Caldwell} (hereafter QP-models).
Now using eq.(\ref{nfe1}) the normalized Hubble parameter becomes
\be
E^{2}(a)=\Omega_{m}a^{-3}+(1-\Omega_{m})a^{-3(1+w)} \;\;.
\ee
Comparing the QP-models with the observational data
(we sample $\Omega_{m} \in [0.1,1]$ and $w \in [-2,-0.4]$ in steps of 0.01)
we find that the best fit values are $\Omega_{m}=0.28\pm 0.02$ and
$w=-1.02\pm 0.06$ with
$\chi_{tot}^{2}(\Omega_{m},w)/dof \simeq 309.2/307$
in very good agreement with the
5 years WMAP data \cite{komatsu08}. Also
Davis et al. \cite{essence} using
the Essence-SNIa+BAO+CMB and a Bayesian statistics found
$\Omega_{m}=0.27 \pm 0.04$, while
Kowalski et al. \cite{Kowal08} utilizing the Union08-SNIa+BAO+CMB obtained
$\Omega_{m}=0.285^{+0.02+0.01}_{-0.02-0.01}$. Obviously, our results
coincide within $1\sigma$ errors. It is worth noting that
the concordance $\Lambda$-cosmology can be described by QP models with
$w$ strictly equal to -1. In this case we find: $\Omega_{m}=0.28\pm 0.02$
with $\chi_{tot}^{2}(\Omega_{m})/dof\simeq 309.3/308$.

\subsection{The Braneworld Gravity - BRG model}
In the context of a braneworld cosmology (hereafter BRG) the accelerated
expansion of the universe can be explained by a modification of gravity
in which gravity itself becomes weak at very large distances (close to the
Hubble scale) due to the fact that our four dimensional brane
survives into an extra dimensional manifold (see \cite{Deff}
and references therein). The interesting point in this scenario is that
the corresponding functional form of the normalized Hubble flow,
eq. (\ref{nfe2}), is affected only by one
free parameter ($\Omega_{m}$). Notice, that
the quantity $\delta H^{2}$ is given by
\be
\delta H^{2}=2\Omega_{bw}+2\sqrt{\Omega_{bw}}
\sqrt{\Omega_{m}a^{-3}+\Omega_{bw}}
\ee
where $\Omega_{bw}=(1-\Omega_{m})^{2}/4$.
The geometrical dark energy equation of state parameter
(see eq.\ref{eos222}) reduces to
\be
w(a)=-\frac{1}{1+\Omega_{m}(a)}
\ee
where $\Omega_{m}(a)\equiv \Omega_{m}a^{-3}/E^{2}(a)$.
The joint likelihood analysis provides a best fit value of
$\Omega_{m}=0.24\pm 0.02$, but the fit is rather poor
$\chi_{tot}^{2}(\Omega_{m})/dof\simeq 369/308$.

\subsection{The parametric Dark Energy - CPL model}
In this model we utilize the  Chevalier-Polarski-Linder
(\cite{Chevallier:2001qy,Linder:2002et}, hereafter CPL)
parametrization. The dark energy equation of state parameter is
defined as a first order Taylor expansion around the present
epoch:
\begin{equation}
w(a)=w_{0}+w_{1}(1-a) \;\;\; .\label{cpldef}
\end{equation}
The normalized Hubble parameter is given by (see eq.\ref{nfe1}):
\begin{equation}
E^{2}(a)=\Omega_{m}a^{-3}+(1-\Omega_{m})
a^{-3(1+w_{0}+w_{1})}e^{3w_{1}(a-1)}
\end{equation}
where $w_{0}$ and $w_{1}$ are constants.
We sample the unknown parameters as follows: $w_{0} \in [-2,-0.4]$
and $w_{1} \in [-2.6,2.6]$.
We find that for the prior of $\Omega_{m}=0.28$ the overall
likelihood function
peaks at $w_{0}=-1.1^{+0.22}_{-0.16}$
and $w_{1}=0.60^{+0.62}_{-1.54}$
while the corresponding $\chi_{tot}^{2}(w_{0},w_{1})/dof$ is 307.6/307.

\subsection{The low Ricci dark energy - LRDE model}
In this modified cosmological model, we use a simple
parametrization for the Ricci scalar which is based on a Taylor
expansion around the present time: ${\cal R}=r_{0}+r_{1}(1-a)$
[for more details see \cite{Linder2004}]. It is interesting to
mention that at the early epochs the cosmic evolution tends
asymptotically to be matter dominated. In this framework, we have
\begin{equation}
E^{2}(a)=\left\{ \begin{array}{cc}
        a^{4(r_{0}+r_{1}-1)}{\rm e}^{4r_{1}(1-a)} & \;\;\;\;a\ge a_{t}  \\
%       \frac{{\rm e}^{mt}-1}{m} & \;\;m\ne 0 \\
%       \mbox{for}\\
       \Omega_{m}a^{-3} & \;\;\;\;a<a_{t}
%\mbox{Homogeneous DE}
       \end{array}
        \right.
\label{SS}
\end{equation}
where $a_{t}=1-(1-4r_{0})/4r_{1}$. The
matter density parameter at the present epoch, is directly related
with the above constants via
\be
\Omega_{m}=\left(\frac{ 4r_{0}-4r_{1}-1 } {4r_{1}}
\right)^{4r_{0}+4r_{1}-1}
{\rm e}^{1-4r_{0}} \;\;.
\ee
The corresponding equation of state parameter is given by
\be
w(a)=\frac{1-4{\cal R}}{3}\left[1-\Omega_{m}
{\rm e}^{-\int_{a}^{1}(1-4{\cal R})(dy/y)}\right]^{-1} \;\;.
\ee
Notice, that we sample the unknown parameters as follows: $r_{0} \in [0.5,1.5]$
and $r_{1} \in [-2.4,-0.1]$ and here are the results:
$r_{0}=0.82 \pm 0.04$
and $r_{1}=-0.74^{+0.10}_{-0.08}$ ($\Omega_{m} \simeq 0.28$) with
$\chi_{tot}^{2}(r_{0},r_{1})/dof \simeq 309.8/307$.

\subsection{The high Ricci dark energy - HRDE model}
A different than the previously mentioned
geometrical method was proposed by Linder \cite{Linder2007},
in which the Ricci scalar at high redshifts evolves as
\be
{\cal R}\simeq \frac{1}{4}\left[1-3w_{1} \frac{\delta H^{2}}
{H^{2}}\right]
\ee
where $\delta H^{2}=E^{2}(a)-\Omega_{m}a^{-3}$.
In this geometrical model the normalized Hubble flow becomes:
\be
E^{2}(a)=\Omega_{m}a^{-3}
\left(1+\beta a^{-3w_{1}}\right)^
{-{\rm ln}\Omega_{m}/{\rm ln}(1+\beta)}
\ee
where $\beta=\Omega_{m}^{-1}-1$.
Using the same sampling as in the QP-models, the joint likelihood
peaks at $\Omega_{m}=0.28\pm 0.03$
and $w_{1}=-1.02\pm 0.1$ with
$\chi_{tot}^{2}(\Omega_{m},w_{1})/dof\simeq 309.2/307$. To this end,
the effective equation of state parameter
is related to $E(a)$ according to eq.(\ref{eos22}).

\subsection{The tension of cosmological magnetic fields  - TCM model}
Recently, \cite{Cont2007} proposed a novel
idea which is based on the following consideration (hereafter TCM): if the
cosmic magnetic field is generated in sources (such as
galaxy clusters) whose overall dimensions
remain unchanged during the expansion of the Universe,
the stretching of this field by the cosmic expansion
generates a tension (negative pressure) that could possibly explain a small
fraction of the dark energy ($\sim 5-10\%$). In this flat
cosmological scenario the normalized Hubble flow becomes:
\be
E^{2}(a)=\Omega_{m}a^{-3}+\Omega_{\Lambda}+\Omega_{B}a^{-3+n}
\ee
where $\Omega_{B}$ is the density parameter for the
cosmic magnetic fields and $\Omega_{\Lambda}=1-\Omega_{m}-\Omega_{B}$.
The equation of state parameter which is related
to magnetic tension is (see eq.\ref{nfe1})
\be
w(a)=-\frac{3\Omega_{\Lambda}+n\Omega_{B}a^{-3+n}}
{3(\Omega_{\Lambda}+\Omega_{B}a^{-3+n})}\;\;.
\ee
To this end, we sample $\Omega_{B} \in [0,0.3]$ and $n \in [0,10]$
and we find that for the prior of $\Omega_{m}=0.28$ the best fit
values are: $\Omega_{B}=0.10\pm 0.10$ and $n=3.60^{+4.5}_{-2.60}$
with $\chi_{tot}^{2}(\Omega_{B},n)/dof\simeq 308.9/307$.

\subsection{The Pseudo-Nambu Goldstone boson - PNGB model}
In this cosmological model the dark energy equation of state
parameter is expressed
with the aid the potential $V(\phi)\propto [1+cos(\phi/F)]$ \cite{Sorbo2007}:
\be
w(a)=-1+(1+w_{0})a^{F}
\ee
where $w_{0} \in [-2,-0.4]$ and $F \in [0,8]$.
Obviously, for $a\ll 1$ we get
$w(a) \longrightarrow -1$.
Based on this parametrization the normalized Hubble function
is given by
\be
E^{2}(a)=\Omega_{m}a^{-3}+(1-\Omega_{m})\rho_{X}(a) \;\;.
\ee
In this context, the corresponding equation of state parameter is
\be
\rho_{X}(a)={\rm exp}\left[\frac{3(1+w_{0})}{F}(1-a^{F}) \right] \;\;.
\ee
Notice, that the likelihood function peaks at $w_{0}=-1.04\pm 0.17$
and $F=5.9\pm 3.2$ with $\chi_{tot}^{2}(w_{0},F)/dof\simeq 317/307$.

\subsection{The early dark energy - EDE model}
Another cosmological scenario that we include in our paper is
the  early dark energy model (hereafter EDE).
The basic assumption here is that at early epochs the amount of
dark energy is not negligible \cite{Doran2006}.
In this framework, the overall dark energy component is given by
\be
\Omega_{X}(a)=\frac{1-\Omega_{m}-\Omega_{e} (1-a^{-3w_{0}})}
{1-\Omega_{m}-\Omega_{m}a^{3w_{0}}}+\Omega_{e}(1-a^{-3w_{0}})
\ee
where $\Omega_{e}$ is the early dark energy density and $w_{0}$
is the equation of state parameter at the present epoch. Notice, that
the EDE model was designed to simultaneously (a) mimic the effects of the
late dark energy and (b) provide a decelerated expansion of the universe at high redshifts.
The normalized Hubble parameter is written as
\be
E^{2}(a)=\frac{\Omega_{m}a^{-3}}{1-\Omega_{X}(a)}
\ee
while using eq.(\ref{eos22}), we can obtain the
equation of state parameter as a function of the scale factor.
Now, from the joint likelihood analysis we find that
$\Omega_{e}=0.05\pm 0.04$ and $w_{0}=-1.14^{+0.18}_{-0.10}$
(for the prior of $\Omega_{m}=0.28$) with
$\chi_{tot}^{2}(\Omega_{e},w_{0})/dof\simeq 308.7/307$.

\subsection{The Variable Chaplygin Gas - VCG model}
Let us consider now a completely different model namely the
variable Chaplygin gas (hereafter VCG) which corresponds
to a Born-Infeld tachyon action \cite{Bento03,Guo05}. Recently, an
interesting family of Chaplygin gas models was found to be consistent
with the current observational data \cite{Vcgdata}.
In the framework of a
spatially flat geometry,
it can be shown that the normalized Hubble function takes the following formula:
\begin{equation}
E^{2}(a)=\Omega_{b}a^{-3}+(1-\Omega_{b})
\sqrt{B_{s}a^{-6}+(1-B_{s})a^{-n}}
\end{equation}
where $\Omega_{b}\simeq 0.021h^{-2}$ is the density
parameter for the baryonic matter \cite{Kirk03} and
$B_{s} \in [0.01,0.51]$ and $n\in [-4,4]$.
The effective matter density parameter is:
$\Omega^{eff}_{m}=\Omega_{b}+(1-\Omega_{b})\sqrt{B_{s}}$.
We find that the best fit parameters are
$B_{s}=0.05\pm 0.02$ and $n=1.58^{+0.35}_{-0.43}$ ($\Omega^{eff}_{m}\simeq 0.26$)
with $\chi_{tot}^{2}(B_{s},n)/dof \simeq 314.7/307$.

\section{Evolution of matter perturbations}
In this section we study the spherical collapse model
by generalizing the basic non-linear equations
which govern the behavior of the matter perturbations
within the framework of the previously described dark energy models.
Also, we compare our predictions with the traditional $\Lambda$ cosmology.
This can help us to understand better the theoretical expectations of
the current dark energy models as well as the variants from the
concordance $\Lambda$ cosmology.

\subsection{The Evolution of the linear growth factor}
The evolution equation
of the growth factor for models where the dark energy
fluid has a vanishing anisotropic stress and the matter fluid is not
coupled to other matter species is given by
(\cite{Linjen03}, \cite{Peeb93}, \cite{Stab06}, \cite{Uzan07}):
\begin{equation}
D^{''}+\frac{3}{2}\left[1-\frac{w(a)}{1+X(a)}\right]\frac{D^{'}}{a}-
\frac{3}{2}\frac{X(a)}{1+X(a)}\frac{D}{a^{2}}=0
\label{deltatime1}
\end{equation}
where
\begin{equation}
X(a)=\frac{\Omega_{m}}{1-\Omega_{m}}{\rm e}^{-3\int_{a}^{1} w(y) d{\rm
    ln}y}=\frac{\Omega_{m}a^{-3}}{\delta H^{2}} \;\;.
\end{equation}
Note, that the prime denotes derivatives with respect
to the scale factor.
%$N={\rm ln}a$ and $\Omega_{m}(a)=\Omega_{m}a^{-3}/E^{2}(a)$.
Useful expressions of the growth factor can be found for the
$\Lambda$CDM cosmology in \cite{Peeb93}, for dark energy models with
$w=const$ in
\cite{Silv94}, \cite{Wang98}, \cite{Bas03}, \cite{Nes08},
for dark energy models with
a time varying equation of state in
\cite{Linca08} and for the
scalar tensor models in \cite{Gann08}. Finally,
in this work, the growth factor
evolution for the current cosmological model is derived by solving
numerically eq. (\ref{deltatime1}).
Note, that the growth factors are normalized to unity
at the present time.
 \begin{figure}[ht]
         \centerline{\includegraphics[width=23pc] {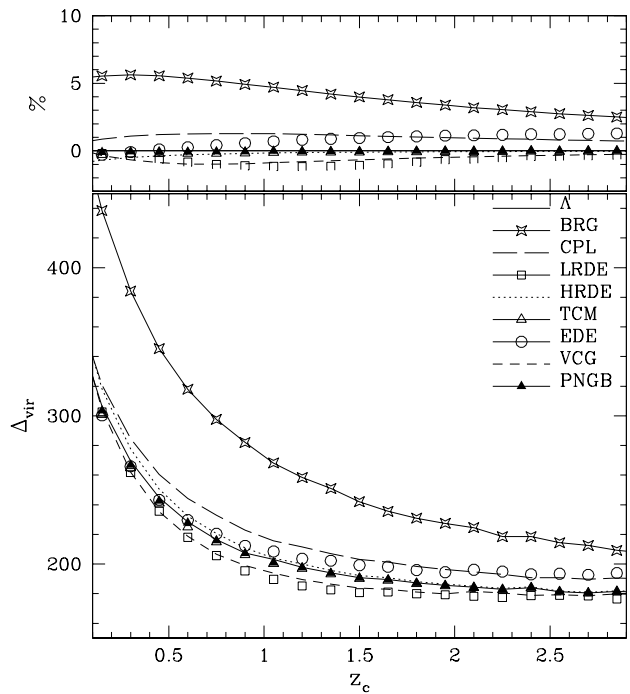}}
  \caption{
{\it Upper Panel:}
The deviation $(1-\lambda_{X}/\lambda_{\Lambda})\%$ of the
collapse factor for various
dark energy models with respect to the $\Lambda$ solution (solid
line).
{\it Bottom Panel:} The density contrast at the virialization,
$\Delta_{vir}$, as a function of redshift.
The points represent the following cosmological models: (a)
BRG (open stars), (b)
LRDE (open squares), (c)
TCM (open triangles), (d)
EDE (open circles) and (e)
PNGB (solid triangles).
The lines represent: (a)
CPL model (long dashed), (b) HRDE model (dot line)
and VCG model (dashed line).
}
        \label{Figcondz}
 \end{figure}

\begin{table}[h]
%\begin{table}[*]
\caption[]{Numerical results. The $1^{st}$ column
indicates the dark energy model used.
Between column two and four, we present the main properties
of the spherical collapse model,
assuming that galaxy clusters have formed prior to
the epoch of $a_{c}\sim 0.42$ ($z_{c}\sim 1.4$), in which
the most distant cluster has been found
\cite{Mul05}. Column five, corresponds to the current age of the universe,
$t_{0}$. Finally, the last two rows correspond
to the $(\gamma_{1},n_{1})$ constants which are included in the approximate
$\Delta_{vir}(a)$ formula (see eq.\ref{aproxf}).}
\tabcolsep 5.8pt
\begin{tabular}{ccccccc} \hline \hline
Model & $z_{ta}$ & $\Delta_{vir}$ &  $\zeta$ & $t_{0}$/Gyr & $\gamma_{1}$ &$n_{1}$\\ \hline
$\Lambda$ & 2.89 & 192.0 & 5.56& 13.72&0&0 \\
QP & 2.89& 191.3 & 5.55 & 13.77&0.384&2.350 \\
BRG & 2.98& 247.3 & 5.89 & 13.40&0.446&-0.556 \\
CPL & 2.90& 204.6 & 5.66 & 13.62&0.210&-1.40 \\
LRDE & 2.86& 181.2 & 5.49 & 13.84&0.152&-6.848 \\
HRDE & 2.89& 193.6 & 5.63 & 13.77&0.050&-0.117 \\
TCM & 2.89& 191.4 & 5.56 & 13.78&0.450&5.387 \\
PNGB& 2.89& 191.9 & 5.57 & 13.76&0&0 \\
EDE& 2.88& 200.1 & 5.71 &  13.71&0.054&-1.70 \\
VCG& 2.85& 185.1 & 5.53 &  14.01&0.226&-1.245 \\
\end{tabular}
\end{table}

\subsection{The spherical collapse model}
The  spherical collapse model, which has a long history
in cosmology, is a simple but still a fundamental tool for
understanding how a small spherical patch [with radius $R(t)$] of
homogeneous overdensity forms a bound system via gravitational
instability \cite{Gunn72}. From now on, we will call $a_{t}$ the
scale factor of the universe where the overdensity reaches its
maximum expansion ($\dot R=0$) and $a_{c}$ the scale factor in
which the sphere virializes implying that a large scale structure
has formed, while $R_{t}$ and $R_{c}$ are the corresponding radii
of the spherical overdensity. Note that in the spherical region,
$\rho_{ms}\propto R^{-3}$ is the matter density, $\rho_{m}$ is the
background matter density and $\rho_{Xs}$ denotes the
corresponding density of the dark energy. In order to address the
issue of how the dark energy itself affects the gravitationally
bound systems (clusters of galaxies), we have to deal with the
following situations: (i) clustered dark energy, considering that
the whole system virializes (both matter and dark energy), (ii)
the dark energy remains clustered but now only the matter
virializes and (iii) the dark energy remains homogeneous and only
the corresponding matter virializes (for more details see
\cite{Mota04}, \cite{Maor05} \cite{Wang06} and \cite{Basi07}).
Note, that in this section we are using the third possibility
(for more on inhomogeneous dark energy see section 7B).

In this section, we review only some basic concepts of the problem
based on the assumption that the dark energy component under a
scale of galaxy clusters can be treated as being homogeneous:
$\rho_{Xs}(a)=\rho_{X}(a)$ and $w_{s}(a)=w(a)$.
In general the evolution of the spherical perturbations as the latter decouple
from the background expansion is given by the Raychaudhuri equation:
\begin{equation}
\frac{\ddot{R}}{R}=-\frac{4\pi G}{3}[\rho_{ms}+\rho_{Xs}(1+3w_{c})]\;\; .
\label{rayc}
\end{equation}
Now, if we perform the transformations
\begin{equation}
x=\frac{a}{a_{t}}\;\;\;\; {\rm and}\;\;\;\;
y=\frac{R}{R_{t}} \;\;\;,
\end{equation}
then eqs.(\ref{fe1}, \ref{rayc}) become:
\begin{equation}\label{xdot}
{\dot x}^{2}=H_{t}^{2}\Omega_{m,t} \left[ \Omega_{m}(x) x
\right]^{-1}
\end{equation}
and
\begin{equation}\label{yddot}
{\ddot y}=-\frac{H_{t}^{2}\Omega_{ m,t}}{2} \left[
\frac{\zeta}{y^{2}}+\nu y I(x,y)\right]
\end{equation}
with
\begin{equation}
\nu=\frac{\rho_{X, t}}{\rho_{m, t}}=\frac{1-\Omega_{m,
    t}}{\Omega_{m, t}}
\end{equation}
where $\rho_{X, t}$ is the dark energy density
at the turn around epoch and
$\Omega_{m,t}$ is the corresponding matter density parameter. Also,
$\rho_{ms, t}=\zeta \rho_{m, t}$ is the matter density
for the spherical region at the turn around time,
while $\rho_{m, t}$ denotes the background matter density at the same epoch.

Note, that in order to obtain the above
set of equations which govern the global and local
dynamics, we have utilized the following relations:
\begin{equation}
\rho_{ms}=\rho_{ms, t} \left(\frac{R}{R_{t}}\right)^{-3}=
\frac{\zeta \rho_{m, t}}{y^{3}}
\end{equation}
and
\begin{equation}
I(x,y)=I(x)=\left[1+3w(x)\right]f(x)
\end{equation}
where
\begin{equation}
f(x)={\rm e}^{3\int_{x}^{1} d{\rm ln}u\left[1+w(u)\right] } \;\;.
\end{equation}
Finally, $\Omega_{m}(x)$ is given by
\begin{equation}
\Omega_{m}(x)=\frac{1}{1+\nu x^{3} f(x)} \;\;.
\end{equation}

In this context, the time needed for a spherical shell to re-collapse
is twice the turn-around time, $t_{f}\simeq 2t_{t}$ which implies
\begin{equation}
\int_{0}^{a_{c}}
\frac{1}{H(a) a} {\rm d}a
=2
\int_{0}^{a_{t}}
\frac{1}{H(a) a} {\rm d}a \;\;.
\label{tim}
\end{equation}
In the case of the usual $\Lambda$ cosmology one can prove that
\begin{equation}
{\rm sinh^{-1}}\left(a_{c}^{3/2} \sqrt{\nu_{0}} \right)=
2{\rm sinh^{-1}}\left(a_{t}^{3/2} \sqrt{\nu_{0}} \right)
\end{equation}
where $\nu_{0}=(1-\Omega_{m})/\Omega_{m}$.
Of course, in order to include the dark energy models in our analysis
we have to solve eq.(\ref{tim}) numerically.
As an example, assuming that galaxy clusters have formed prior to
the epoch of $a_{c}\sim 0.42$ ($z_{c}\sim 1.4$), in which
the most distant cluster has been found
\cite{Mul05}, the turn around epoch
is not really affected by the dark energy component (see Table 1), ie.
$a_{t} \sim 0.26$ (or $z_{t}\sim 2.8$). This is to be expected,
due to the fact that at large redshifts matter dominates the Hubble expansion.
It is worth noting that the ratio between the
scale factors converges to the Einstein de Sitter ($\Omega_{m}=1$) value
$\left(\frac{a_{c}}{a_{t}}\right)=2^{2/3}$
at high redshifts which implies that
$\zeta \simeq \left(\frac{3 \pi}{4}\right)^{2}$.

On the other hand, utilizing both the virial theorem and the energy
conservation\footnote{In fact, a smooth dark energy component violates the energy
conservation, in the sense that the dark energy freely flows outwards the
overdensity \cite{Mota04,Wang06}. Only small scale clustering dark
energy would satisfy the energy conservation, for instance the
chameleon models \cite{cham1,cham2,cham3,cham4}. Such violation is however
only important at very late times, when dark energy dominates.} we reach to the following condition:
\begin{equation}
\left[\frac{1}{2}R\frac{\partial}{\partial R} (U_{G}+U_{Xs})+
U_{G}+U_{Xs}\right]^{a=a_{c}}=
\left[U_{G}+U_{Xs} \right]^{a=a_{t}}
\label{virial}
\end{equation}
where $U_{G}=-3GM^{2}/5R$ is the potential energy and
$U_{Xs}=-4\pi GM(1+3w)\rho_{Xs}R^{2}/5$ is the potential energy
associated with the dark energy for the spherical overdensity (see
\cite{Mota04} and \cite{Maor05}). Using the above formulation we
can obtain a cubic equation that relates the ratio between the
virial $R_{c}$ and the turn-around outer radius $R_{t}$ the  collapse factor ($\lambda=R_{c}/R_{t}$).
%We would like to stress here that
%eq.(\ref{virial}) is valid when the ratio of the system's dark energy
%to the matter's densities at the time of the
%turn-around takes relatively small values \cite{Wang06}.
Of course in the case of $w=-1$ the above expressions
get the usual form for $\Lambda$ cosmology (\cite{Bas03}, \cite{Lahav91})
while for an Einstein-de Sitter model ($\Omega_{m}=1$)
we have $\lambda=1/2$.

Finally solving numerically eq.(\ref{virial})
we calculate the collapse factor. In general,
we find that the collapse factor
lies in the range $0.43\le \lambda \le 0.50$ in agreement with
previous studies (
\cite{Bas03}, \cite{Mota04}, \cite{Maor05}, \cite{Wang06}, \cite{Horel05},
\cite{Perc05}, \cite{Basi07}).
Also, in the upper panel of figure 1 we
plot the deviation, $(1-\lambda_{X}/\lambda_{\Lambda})\%$,
of the collapse factors
$\lambda_{X}(z_{c})$ for the current dark energy models
with respect to the $\Lambda$ solution $\lambda_{\Lambda}(z_{c})$.
It becomes evident that the shape of the cosmic structures
which produced by the CPL (long dashed line),
LRDE (open squares), HRDE (dot line), TCM (open triangles)
and PNGB (solid triangles) models is close to
that predicted by the usual $\Lambda$ cosmology.
On the other hand, the
largest positive deviation of the collapse factor occurs
for the BRG model (open stars) which implies that
the latter model produces more bound systems with respect to
the above dark energy models. Therefore, within this cosmological
scenario the corresponding large scale structures should locate in very large
density environments. The opposite situation
holds for the VCG (dashed line)
and LRDE (open squares) models due to their negative deviations.

In the bottom panel of figure 1 we present the evolution of the
density contrast at virialization
\begin{equation}\label{deltavir}
\Delta_{vir}=\frac{\rho_{ms, c}}{\rho_{m, c}}=
\frac{\zeta}{\lambda^{3}} \left(\frac{a_{c}}{a_{t}}\right)^{3}
\end{equation}
where $\rho_{ms, c}$ is the matter density in the virialized structure
while $\rho_{m, c}$ is the background matter density
at the same epoch.
We verify, that the density contrast
decreases with
the formation (virialization) redshift $z_{c}$.
Obviously, the factor $\Delta_{vir}$,
provided by the spherical collapse model, plays a
key role in this kind of studies.
Indeed, the differences among the $\Delta_{vir}$ for the
current dark energy models enter
through the $H(z)$ (see eq. (\ref{tim})).
This feature implies that perhaps
the density contrast at virialization can be used as a
cosmological tool (see section 7). In any case, at very large redshifts the
above density contrast tends to the Einstein-de Sitter value
($\Delta_{vir}\sim 18\pi^{2}$), as it should.
In this context, following the notations of
\cite{Wein03, Kita96,shaw}, we provide an accurate fitting
formula to $\Delta_{vir}$
(within a physical range of cosmological parameters)
\be
\label{aproxf}
\Delta_{vir}=18\pi^{2}[1+\gamma \Theta^{n}(a)]
\ee
where
$$\gamma=0.44-1.31(|w(a_{\star})|^{\gamma_{1}}-1)$$
$$n=0.94-0.21(|w(a_{\star})|^{n_{1}}-1)$$
$a_{\star}=0.4$ (or $z_{\star}=1.5$)
and $\Theta(a)=\Omega^{-1}_{m}(a)-1$. Note, that
$\gamma_{1}$ and $n_{1}$ are constants (see the last rows of Table1).

Finally, in Table 1 we list the results of the spherical collapse model
considering that galaxy clusters have formed at $z_{c}\sim 1.4$, ie., (a) the
cosmological models and the value of the turn around redshift, (b) the
virial density $\Delta_{vir}$, at the collapse time, as well as the
density excess of the matter density in the spherical overdensity, $\zeta$,
at the turn around time, (c) the predicted age of the universe and (d)
the constants $(\gamma_{1},n_{1})$ which are included in eq.(\ref{aproxf}).

\section{The formation of galaxy clusters}
In this section we briefly investigate the cluster formation
processes by generalizing the basic equations
which govern the behavior of the matter perturbations
within the framework of the current dark energy models.
Also we compare our predictions with those found for the traditional
$\Lambda$ cosmology. This can help us
to understand better the theoretical expectations of
the dark energy models as well as the variants from
the usual $\Lambda$ cosmology.

\begin{figure}[ht]
\mbox{\epsfxsize=10cm \epsffile{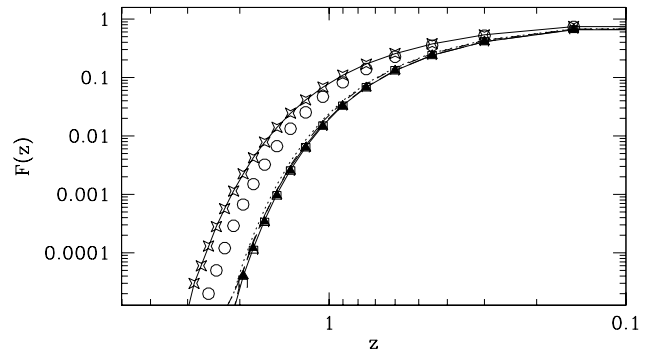}}
\caption{The predicted fractional rate of cluster formation as a
function of redshift for the current cosmological models
($\sigma_{8}=0.80$).}
\end{figure}

The concept of estimating the fractional rate of cluster formation
has been brought up by different
authors \cite{Peeb84, Rich92}. In particular, these
authors introduced a methodology which computes the
rate at which mass joins virialized structures, which grow from small
initial perturbations in the universe.
In particular, the basic tool is the Press-Schechter formalism \cite{press}
which considers the fraction of mass in the universe contained
in gravitationally bound structures (such as galaxy clusters)
with matter fluctuations greater than a critical value $\delta_{c}$,
\cite{eke}.
Assuming that the density contrast is normally distributed
with zero mean and variance $\sigma^{2}(M,z)$
we have:
\be\label{eq:88}
{\cal P}(\delta,z)=\frac{1}{\sqrt{2\pi}\sigma(M,z)}
{\rm exp}\left[-\frac{\delta^{2}}{2\sigma^{2}(M,z)} \right] \;\;.
\ee
In this kind of studies it is common to
parametrize the rms mass fluctuation amplitude at
8 $h^{-1}$Mpc which can be expressed as a function of redshift as
$\sigma(M,z)=\sigma_{8}(z)=D(z)\sigma_{8}$.
The current cosmological models are normalized by
the analysis of the WMAP 5 years data
$\sigma_{8}=0.80$ \cite{komatsu08}.
The integration of eq.(\ref{eq:88}) provides the fraction of the universe,
on some specific mass scale, that has already
collapsed producing cosmic structures (galaxy clusters)
at redshift $z$ and is given by \cite{Rich92}:
\be
%\label{eq:89}
P(z)=\int_{\delta_{c}}^{\infty} {\cal P}(\delta, z) d\delta
\ee
or
\be
\label{eq:89}
P(z)=
\frac{1}{2} \left[1-{\rm erf}
\left( \frac{\delta_{c}}{\sqrt{2} \sigma_{8}(z)} \right) \right]
\ee
where $\delta_{c}$ is the linearly extrapolated density
threshold above which structures collapse, \cite{eke}.
Notice, that for the model of modified gravity (BRG) we use
$\delta_{c}\simeq 1.47$ (see \cite{Shaef08}), for the EDE model
we use $\delta_{c}\simeq 1.4$ (see \cite{Fran08} and references therein).
To this end, for the rest of the dark energy models,
due to the fact that $w \simeq -1$ close to the present epoch,
we utilize a $\delta_{c}$ approximation which is given by
Weinberg \& Kamionkowski (see \cite{Wein03}, their eq.18).
%In this kind of studies it is common to
%parametrize the rms mass fluctuation amplitude at
%8 $h^{-1}$Mpc which can be expressed as a function of redshift as
%$\sigma(M,z)=\sigma_{8}(z)=D(z)\sigma_{8}$.
%The current cosmological models are normalized by
%the analysis of the WMAP 5 years data
%$\sigma_{8}=0.80$ \cite{komatsu08}.

Obviously the above generic of form eq.(\ref{eq:89})
depends on the choice of the background cosmology.
The next step is to normalize the probability to give the number of clusters which
have already collapsed by the epoch $z$ (cumulative distribution), divided
by the number of clusters which have collapsed at the
present epoch ($z=0$), $F(z)=P(z)/P(0)$.
In figure 2 we present in a logarithmic scale the behavior of normalized
cluster formation rate as a function of redshift for the
present dark energy models. In general, prior to $z\sim 0.2$ the cluster
formation has terminated
due to the fact that the matter fluctuation field, $D(z)$, effectively freezes.
For the traditional
$\Lambda$ cosmology we find the known behavior in
which galaxy clusters appear to be formed at high redshifts $z\sim 2$ (see
for example \cite{Bas03} and references therein).
From figure 2 it becomes also clear, that the vast majority of the dark energy models
seem to have a cluster formation rate which is close to that predicted
by the usual $\Lambda$ cosmology (see solid line in figure 2).
However, for the BRG and EDE cosmological scenarios respectively we find that galaxy
clusters appear to form earlier ($z\sim 3$) with respect
to the CPL, LRDE, HRDE, TCM, PNGB and VCG
dark energy models.

\section{Comparison with Observations}
\subsection{Homogeneous Dark energy}
A useful observable parameter that is predictable in the context
of the spherical collapse model is the concentration parameter
$c_{vir}$. This parameter characterizes the profile of cluster
dark matter halo which in turn is usually modelled by the NFW
profile \cite{Navarro:1995iw,Navarro:1996gj} defined as \be
\rho_{NFW}(R) = \frac{\rho_s}{(R/r_s)(1 + R/r_s)^2} \label{nfw}
\ee where $r_s$ and $\rho_s$ are the characteristic radius and
density of the halo profile. The concentration parameter $c_{vir}$
is an observable characteristic of the halo profile and is defined
as $c_{vir}=r_{vir}/r_s$ where $r_{vir}$ is the virial radius of
the system \cite{Navarro:1995iw,Navarro:1996gj,Wechsler:2001cs}.
Dark matter profiles of clusters may be determined using lens
distortion and magnification
\cite{Broadhurst:2008re,Umetsu:2007pq}. Such profiles have been
recently obtained \cite{Broadhurst:2008re} for four clusters using
the wide-field camera Suprime-Cam \cite{Miyazaki:2002wa} of the
Subaru 8.3m telescope. The NFW profile fits well the profile of
these clusters and the concentration parameters have been obtained
for all four clusters \cite{Broadhurst:2008re}.
\begin{figure*}[ht]
\centering
\begin{center}
$\begin{array}{@{\hspace{-0.1in}}c@{\hspace{0.0in}}c}
\multicolumn{1}{l}{\mbox{}} &
\multicolumn{1}{l}{\mbox{}} \\ [-0.20in]
\epsfxsize=3.3in
\epsffile{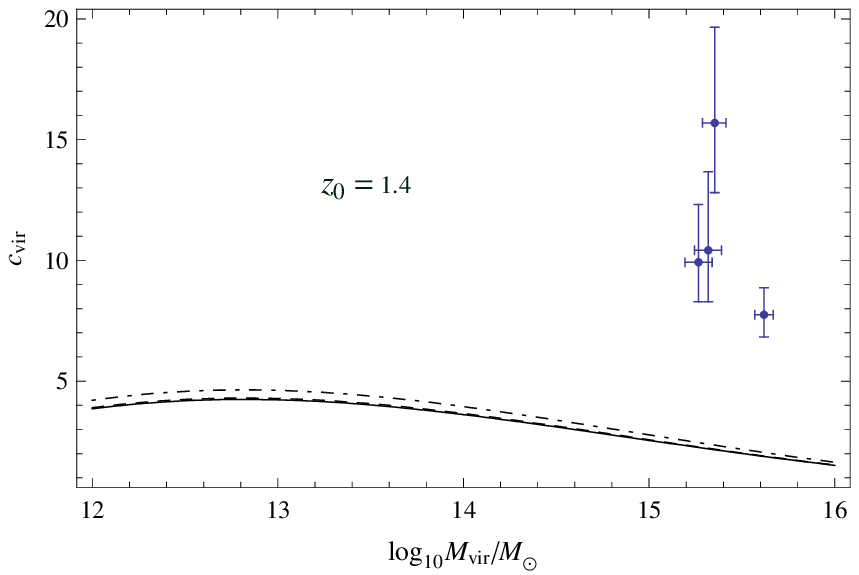} &
\epsfxsize=3.3in
\epsffile{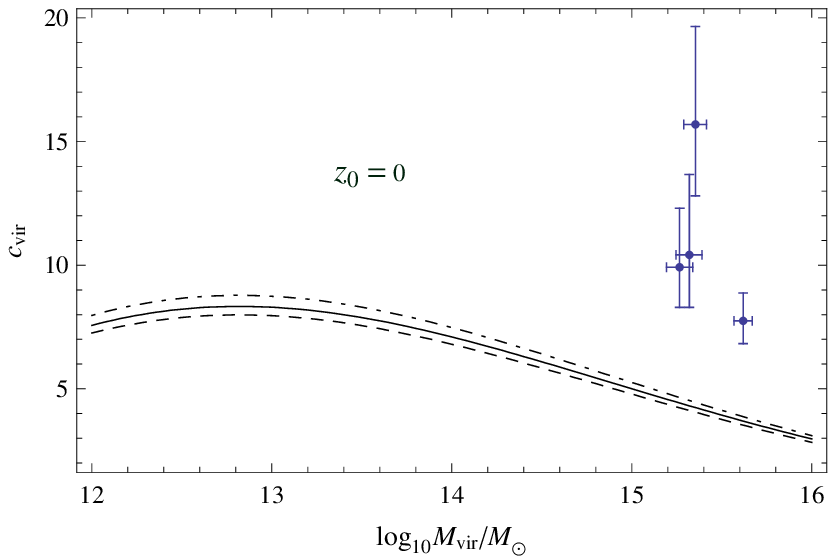} \\
\end{array}$
\end{center}
\vspace{0.0cm} \caption{\small a: The concentration parameter
$c_{vir}(M_{vir})$ for homogeneous dark energy for
$(w_0,w_1)=(-1,0)$ (\lcdm) (solid line), $(w_0,w_1)=(-1.1,0.6)$ (best fit of section 4C) and $(w_0,w_1)=(-0.6,0)$
(dot-dashed line). For comparison we include the predicted curve
(dashed line) for the values at which the overall likelihood
function peaks: $(w_0,w_1)=(-1.1,0.6)$ (see
Sec.~\ref{sec.constr}). Redshift $z_0=1.4$ has been
assumed. b: Same as Fig. 3a with $z_0=0$.} \label{fig3}
\end{figure*}

The theoretical prediction of the concentration parameter for each
cluster can be made using the results of the previous section in
the context of the spherical collapse model with any dark energy
parametrization assuming homogeneity of dark energy. In what
follows we consider the CPL parametrization of equation
(\ref{cpldef}). In particular, we use the approach of Ref.
\cite{Eke:2000av} (see also \cite{Mota:2008ne}) to obtain the
predicted value of the concentration parameter from the density
contrast at virialization $\Delta_{vir}$. This is a simple
effective algorithm that approximates well the values of the
concentration parameter obtained from N-body simulations in the
context of various cosmological dark energy models including the
case of the cosmological constant. In this approach, the
concentration parameter $c_{vir}$ is related to $\Delta_{vir}(z)$
as\footnote{Note that $\Delta_{vir}$ entering Eq.~(\ref{cvireval})
is defined with respect to the critical density $\rho_{crit}$
\cite{Eke:2000av}, rather than with respect to $\rho_{m,c}$ as in
Eq.~(\ref{deltavir}). Computing $c_{vir}$ then requires to
multiply $\Delta_{vir}$ in Eqs.~(\ref{deltavir}) and
(\ref{aproxf}) by $\rho_{m,c}/\rho_{\rm crit}$.} \be c_{vir}^3
=\frac{\Delta_{vir}(z_c) \; \Omega_m(z_0)
(1+z_c)^3}{\Delta_{vir}(z_0) \; \Omega_m(z_c) (1+z_0)^3}
\label{cvireval} \ee where $z_c$ is the formation (virialization)
redshift while $z_0$ is the ``observed'' redshift ($z=z_{0} \leq
z_{c}$). The formation redshift $z_c$ may be obtained
\cite{Eke:2000av} from the relation \be D(z_c)
\sigma_{eff}(M)=\frac{1}{C_\sigma} \label{zceval} \ee where $D(z)$
is the linear growth factor and $\sigma_{eff}(M)$ is the {\it
effective} amplitude of the power spectrum on a scale $M$
connected with the actual amplitude on a scale $M$ by \be
\sigma_{eff}(M)=\sigma(M)\left[-\frac{d\ln \sigma}{d\ln M}
(M)\right] \label{sigeff} \ee $C_{\sigma}$ is a constant with
$C_\sigma \simeq 28$ \cite{Eke:2000av}. The power spectrum
amplitude $\sigma(M)$ is well approximated by the fit
\cite{Liberato:2006un} \be \sigma(M)=\sigma_8
\left(\frac{M}{M_8}\right)^{-\gamma_{1} (M)/3} \label{sigm1} \ee
where $M_8=6\times 10^{14} \Omega_{0m}h^{-1} M_{\odot}$ is the
mass inside a sphere of radius $R_8=8h^{-1}$Mpc.
%and $\sigma_8$
%is the variance of the overdensity field smoothed on a scale of size $R_8$.
Also
\be \gamma_{1}(M)= (0.3 \Gamma + 0.2)[2.92 + \frac{1}{3} \log(\frac{M}{M_8})] \label{gamma-m}
\ee
with $\Gamma=\Omega_{0m} h e^{-\Omega_b-\Omega_b/\Omega_{0m}}$
($\Omega_b=0.05$ is the baryonic density parameter).
The linear growth function
in equation (\ref{zceval}) is
obtained as (eg \cite{Nesseris:2007pa})
\be
D(z)=e^{\int_1^{1/(1+z)}\Omega_m(a)^\gamma \frac{da}{a}} \label{grfac}
\ee where $\gamma$ is the effective growth index which depends mildly on the
dark energy properties in the observationally allowed parameter range ($\gamma=0.55$ for \lcdm).

Using equations (\ref{zceval}), (\ref{sigm1}), (\ref{gamma-m}) and
the parametrization (\ref{aproxf}) to evaluate $\Delta_{vir}(z)$
we may use equation (\ref{cvireval}) to evaluate
$c_{vir}(M_{vir})$ for various values of redshift $z_0$ and
compare with the corresponding data of Ref.
\cite{Broadhurst:2008re}. The results are shown in Fig. 3a for
$z_0=1.4$ and in Fig. 3b for $z_0=0$. In both cases we consider
$(w_0,w_1)=(-1,0)$ (\lcdm) and $(w_0,w_1)=(-0.6,0)$ which slightly
improves the fit to the data of Ref. \cite{Broadhurst:2008re}
shown in the same figures (values $w_0<-1$ further decrease the
quality of fit). We have also tested other cases with $w_1\neq 0$ and we
have found that the value of $c_{vir}$ depends (mildly) only on
the {\it mean value} of $w(z)$ at late times when dark energy
begins to dominate.

The results shown in Fig. 3 correspond to homogeneous dark energy
and confirm the well known puzzle for \lcdm$ $  namely that the
predicted concentration parameter in this model in significantly
less than the observed one \cite{Broadhurst:2008re}. The new input
of Fig. 3 is that the modification of dark energy evolution
(within the observationally allowed range) is unable to resolve
this puzzle if homogeneity is assumed for dark energy. As
discussed in the next subsection however the agreement with
observations can improve significantly if dark energy is assumed
to cluster along with the non-linearly clustered dark matter.

\subsection{Inhomogeneous Dark energy}
Although it has been shown that the dark energy fluid does not
cluster on scales smaller than 100Mpc \cite{Dave:2002mn}, it is
also interesting to consider the case when the dark energy
clusters along with the dark matter. Such a behavior is expected in models of interacting dark energy with dark matter and may also be a result of the gravitational interaction between nonlinearly clustered dark matter and dark energy. In the homogeneous case, the dark energy component
flows progressively out of the overdensity \cite{Maor05,david},
and hence energy conservation cannot be applied to determine the
collapse factor $\lambda$ (along with the virial theorem). In
order to simplify the inhomogeneous formalism, we consider the
extreme case in which the dark energy fully clusters along with
the dark matter and thus we avoid the energy non-conservation
problem examined in Ref.~\cite{Maor05}.
%It is convenient, though, to
%emphasize that yet another energy non-conservation problem arises
%in QCDM models, as owing to $w\neq-1$ the dark energy contributes
%a non-conservative force. Nevertheless, the repercussion of this
%non-conservation on the collapse factor is negligible unless the
%dark energy contributes significantly to the total energy in the
%cluster at the time of turn-around \cite{Wang06}.
The merit of this assumption is that
it allows an analytical solution to the system of
Eqs.~(\ref{xdot}) and (\ref{yddot}) \cite{Basi07} due to the fact that
the function $I(x,y)$ in eq.(\ref{yddot}) depends only on $y$
\be
I(x,y)=I(y)=\left[1+3w(R(y))\right]\frac{f(R(y))}{f(\alpha_{\rm t})}  \;\;.
\ee
Therefore, the solution of the system is
\begin{eqnarray}
\label{inHomsol}
\int_0^1\left[\frac{y}{\zeta+\nu y
P(y)-(\zeta+P(1)\nu)y}\right]^{1/2}dy
\nonumber\\
=\int_0^1\left[x\Omega_m(x)\right]^{1/2}dx
\end{eqnarray}
where $P(y)=y^2f(R(y))/f(a_t)$ and $R(y)=\zeta^{-1/3}a_ty$.
The matter contrast $\zeta$ can be obtained by solving numerically
the above equation.

\begin{figure*}[ht]
\centering
\begin{center}
$\begin{array}{@{\hspace{-0.1in}}c@{\hspace{0.0in}}c}
\multicolumn{1}{l}{\mbox{}} & \multicolumn{1}{l}{\mbox{}} \\
[-0.20in] \epsfxsize=3.3in \epsffile{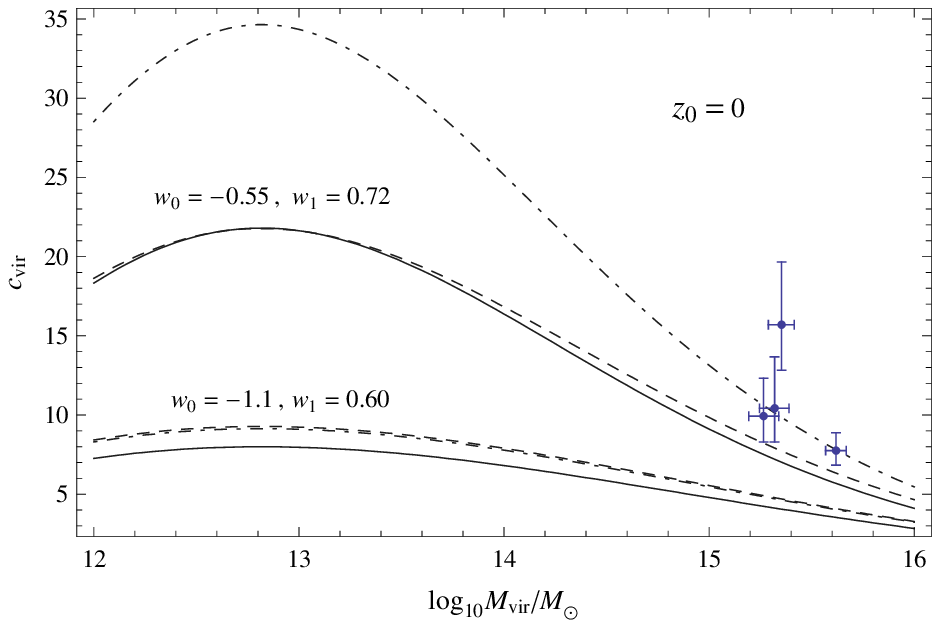} & \epsfxsize=3.3in
\epsffile{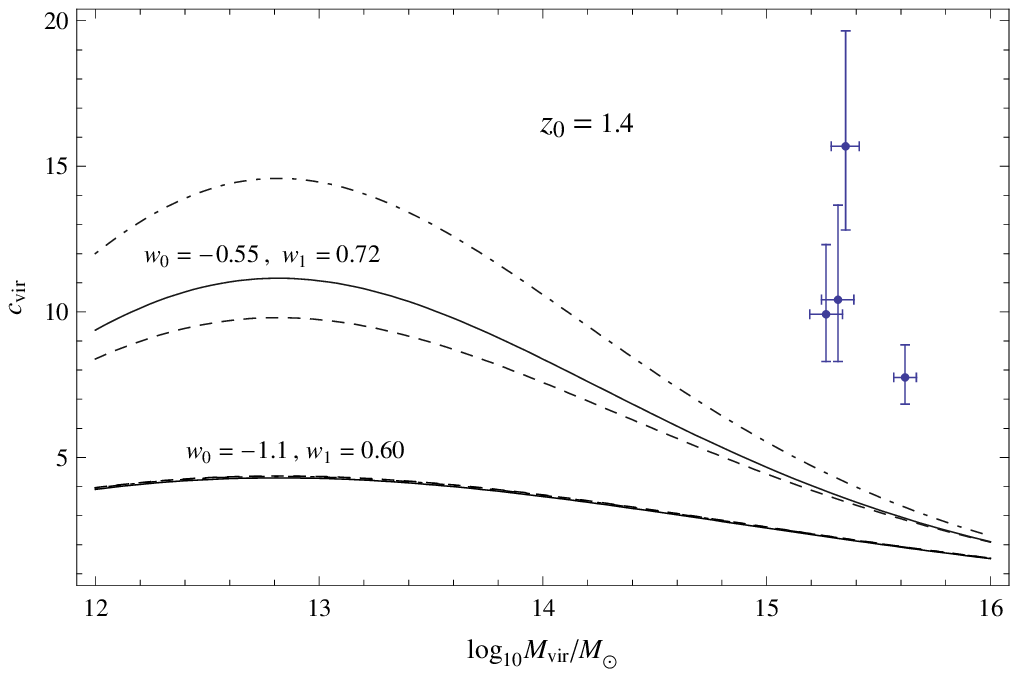} \\
\end{array}$
\end{center}
\vspace{0.0cm} \caption{\small a: Concentration parameter
$c_{vir}(M_{vir})$ for the inhomogeneous dark energy when only
matter virializes (dashed line) and when both matter and
quintessence virialize (dot-dashed line) with $z_0=0$. We include
the homogeneous case (solid line) for comparison. The three upper
curves correspond to $(w_0,w_1)=(-0.55,0.72)$, whereas the three
bottom ones correspond to the values where the overall likelihood
function peaks: $(w_0,w_1)=(-1.1,0.60)$. b: Same as Fig. 4a with
$z_0=1.4$.} \label{fig4}
\end{figure*}

Although the dark energy clusters, it may not take part in the
virialization process, in which case the collapse factor is
determined by eq.~(\ref{virial}). However, if the dark energy
participates in the virialization, the self-energy of the dark
energy must also be taken into account. In such a case, the
equation that determines the collapse factor is \cite{Maor05}
\begin{equation}
\left[\frac{1}{2}R\frac{\partial U_{\rm tot}}{\partial R}+ U_{\rm
tot}\right]^{a=a_{c}}= \left[U_{\rm tot}\right]^{a=a_{t}}\,,
\label{virial2}
\end{equation}
where
\[
U_{\rm tot}=\frac12\int(\rho_m+\rho_{Xs})(\Phi_m+\Phi_{Xs})dV\,,
\]
and $\Phi_{m}$ and $\Phi_{Xs}$ are the gravitational potentials
of the matter and dark energy components for a spherical
overdensity.

After solving for $\zeta$ and $\lambda$ we use eq.~(\ref{tim}) to
compute $\Delta_{vir}$ from eq.~(\ref{deltavir}) at different
redshifts, and then $c_{vir}$ from eq.~(\ref{cvireval}). It was
noted in Ref.~\cite{david} that the $c_{vir}$ corresponding to
inhomogeneous dark energy models may be larger than the
corresponding to their homogeneous realizations, although such
enhancement in $c_{vir}$ is model dependent. Such an increase in
$c_{vir}$ owes primarily to the factor
$\Delta_{vir}(z_c)/\Delta_{vir}(z_0)$ in eq.~(\ref{cvireval}), and
it is shown in \cite{david} that for certain models this factor
may be significantly larger than 1. For example, the
inhomogeneous realization of the model in Ref.~\cite{Brax:1999gp}
results in a $c_{vir}$ that can be larger by a factor of 2.

In the CPL model we find that, in general, although $c_{vir}$ may
be somewhat larger in the inhomogeneous case, this improvement is
insufficient to account for observations if the dark energy
component is sub-dominant at the time of turn-around. The
situation improves when we allow the dark energy to give a
non-negligible contribution at turn-around, which can be arranged
by taking $w_1\gtrsim w_0$ (see Fig~\ref{fig4}). As mentioned
before, in such a case energy conservation cannot be used to
determine the collapse factor. However, it is interesting to
examine the predictions for $c_{vir}$ obtained by pretending that
energy is conserved, hence using eqs.~(\ref{virial}) and
(\ref{virial2}) to determine $\lambda$. Of course, in this case
one must understand $c_{vir}$ as an estimate to its actual value.

Assuming then energy conservation to estimate $\lambda$, we begin
by the case when only the matter component virializes. In that
case $\Delta_{vir}(z_c)/\Delta_{vir}(z_0)$ in eq.~(\ref{cvireval})
is of order 1, and so the enhancement factor in $c_{vir}$ is lost.
We find that in the range of $M_{vir}$ of interest to
observations: $10^{15}M_\odot\lesssim M_{vir}\lesssim
10^{16}M_\odot$ \cite{Broadhurst:2008re}, the concentration
parameter behaves roughly as $c_{vir}\sim \frac{1+z_c}{1+z_0}$.
The agreement with observations then depends almost exclusively on
$z_0$ and $z_c$, requiring an early rapid collapse of the halo
subunits: $z_c\sim7$ for $z_0\simeq0$. The agreement with
observations ameliorates if the dark energy component also
virializes. The reason is that in such a case the corresponding
collapse factor is smaller than if only matter virializes
\cite{Basi07}. Consequently, the factor
$\Delta_{vir}(z_c)/\Delta_{vir}(z_0)$ increases, becoming
significantly larger than 1 in the observational range of
$M_{vir}$ provided $w_1$ is properly tuned (see Fig.~\ref{fig4}).
Similar results to the ones presented in Fig.~\ref{fig4} may be
obtained for a wide range of values of $w_0$ (assuming of course
$w_1$ is conveniently tuned), in particular for $w_0\simeq-1$,
which is in better agreement with current observations. In
Fig.\ref{fig4}.b we present the results for $c_{vir}$ for
$z_0=1.4$. In this case, the factor
$\Delta_{vir}(z_c)/\Delta_{vir}(z_0)$ is counteracted by the term
$\frac{1+z_c}{1+z_0}$, and hence agreement with observations
cannot be attained in this case.

To conclude we would like to emphasize that the results presented
in Fig.~\ref{fig4}, owing to the aforementioned energy
non-conservation problem, must be understood as an estimate to the
actual value of $c_{vir}$. A more reliable estimate for $c_{vir}$
must obviously account for energy non-conservation in the equation
determining the collapse factor $\lambda$. As pointed out in
\cite{Wang06}, taking this effect into account results in a
further reduction of the collapse factor $\lambda$. Therefore, one
can expect that a more detailed study will place $c_{vir}$ in
better agreement with observations.

\section{Conclusions}
In this paper we study analytically and numerically the spherical
collapse model beyond the usual $\Lambda$ cosmology, by using
several parameterizations for the dark energy equation of state.
In this framework, we first perform a joint likelihood analysis in
order to put tight constraints on the main cosmological parameters
by using the current observational data (SNIa, CMB shift parameter
and BAOs). For the vast majority of the dark energy models, we
find that the large scale structures (such as galaxy clusters)
start to form prior to $z\sim 1-2$ \cite{Rich92,Basi07}. The amplitude and the shape of
the concentration parameter ($c_{vir}$) is affected by the
considered status of the dark energy model (homogeneous or
clustered). We verify that in the case where the distribution of
the dark energy is clustered we are producing more concentrated
structures with respect to the
homogeneous dark energy. Finally, we go a step further by
comparing the predicted concentration parameters with those
observed for four massive galaxy clusters ($0.18\le z \le 0.45$)
and we find that the homogeneous dark energy pattern is unable to
reproduce the data. The situation becomes somewhat better in the case
of an inhomogeneous (clustered) dark energy. In order to confirm
such a possibility, we need to create robust cluster surveys at
relatively high redshifts ($1.5\ge z \ge 0.2$). The latter result
points to the direction that perhaps the $c_{vir}$ parameter
can be used in the future in order to understand the physical
properties of the dark energy.

\acknowledgments
We would like to thank David Mota for his useful comments and suggestions. This work was supported by the European Research and Training
Network MRTPN-CT-2006 035863-1 (UniverseNet).

%Now within the cluster region the evolution of the dark energy
%component is written as (see \cite{Mota04})
%\begin{equation}
%\dot{\rho}_{\phi_{c}}+3\frac{\dot R}{R}(1+w_{\phi_{c}})\rho_{\phi_{c}}
%=\Gamma
%\end{equation}
%while if we consider a scalar field the above equation becomes
%\begin{equation}
%      \ddot{\phi}_{c}+3 \frac{\dot{R}}{R}\dot{\phi}_{c}+U^{\prime}(\phi_{c})=
%\frac{\Gamma}{\dot \phi}
%\end{equation}
%where
%\begin{equation}
%\Gamma=-3\left(\frac{\dot{a}}{a}-\frac{\dot{R}}{R} \right)
%\dot{\phi}^{2}_{c}\;\;\;.
%\end{equation}
%Figure 6 presents examples of $R(t)$ obtained for the UDM (solid line)
%and for the concordance $\Lambda$ model (dashed line).


\begin{thebibliography}{plain}

 \bibitem {Riess07}
 A. G. Riess, {\em et al.}, Astrophys. J., {\bf 659}, 98, (2007)

\bibitem{essence} W.M. Wood-Vasey {\em et al.},
 Astrophys. J., {\bf 666}, 694, (2007);
T.M. Davis {\em et al.}, Astrophys. J., {\bf 666}, 716, (2007)

\bibitem {Spergel07}
 D.N. Spergel, et al., Astrophys. J. Suplem., {\bf 170}, 377, (2007)

\bibitem {komatsu08}
 E. Komatsu, et al., Astrophys. J. Suplem., {\bf 180}, 330, 2009


\bibitem{Kowal08}
M. Kowalski, et al., Astrophys. J., {\bf 686}, 749, (2008)

%4
 \bibitem {Ratra88}
B. Ratra, P. J. Peebles, Phys. Rev D., {\bf 37}, 3406, (1988)

\bibitem{Oze87}
M. Ozer M. and O. Taha, Nucl. Phys., {\bf B287}, 776, (1987)

\bibitem {Weinberg89}
 S. Weinberg, Rev. Mod. Phys., {\bf 61}, 1, (1989)

%5
\bibitem{Wetterich:1994bg}
C.~Wetterich, Astron. Astrophys. {\bf 301}, 321 (1995)

%6
\bibitem{Caldwell98}
R. R. Caldwell, R. Dave, P. J. Steinhardt, Phys. Rev. Lett.,
{\bf 80}, 1582, (1998)

\bibitem{Brax:1999gp}
P.~Brax, J.~Martin, Phys. Lett. {\bf B468}, 40 (1999)

%7
\bibitem{KAM}
A. Kamenshchik, U. Moschella, V. Pasquier, Phys. Lett. B., {\bf 511}, 265, (2001)

\bibitem{fein02}
A. Feinstein, Phys. Rev. D., {\bf 66}, 063511, (2002)
%8
\bibitem{Caldwell}
R.R., Caldwell,  E. V., Phys. Rev. Lett., {\bf 545}, 23, (2002)
%9
\bibitem{Peebles03}
P. J. Peebles, B. Ratra, Rev. Mod. Phys., {\bf 75}, 559, (2003)

\bibitem{chime04}
L. P. Chimento, A. Feinstein, Mod. Phys. Lett. A, {\bf 19}, 761, (2004)

%13
\bibitem{Brookfield:2005td}
A.~W. Brookfield, C.~van~de Bruck, D.~F. Mota, D.~Tocchini-Valentini,
Phys. Rev. Lett. {\bf 96}, 061301 (2006)

%14
\bibitem{Copel06}
 E. J. Copeland, M. Sami, S. Tsujikawa, Int. J. Mod. Phys. D,
{\bf 15}, 1753 (2006)

%15
\bibitem{Boehmer:2007qa}
C.~G. Boehmer, T.~Harko, Eur. Phys. J. {\bf C50}, 423 (2007)

%16
\bibitem{Friem08}
J. Frieman, M. Turner, D. Huterer, Annual Rev. of Astron. and Astrop.,
{\bf 46}, 385, (2008)


\bibitem{Lahav91}O. Lahav, P. B. Lilje, J. R. Primack, \&, M. J. Rees,
Mon. Not. Roy. Astron. Soc., {\bf 251}, 128, (1991)
\bibitem{Wang98}
L.  Wang, \& J. P. Steinhardt, Astrophys. J., {\bf 508}, 483, (1998)
\bibitem{Iliev01}
I. T. Iliev, \&, P. R. Shapiro, Mon. Not. Roy. Astron. Soc., {\bf 325}, 468, (2001)
\bibitem{Battye03}
R. A. Battye, \&, J. Weller, Phys. Rev. D, {\bf 68}, 3506, (2003)

\bibitem{Bas03}
S. Basilakos, Astrophys. J., {\bf 590}, 636, (2003)
\bibitem{manera} M.Manera and D. F. Mota, Mon. Not. Roy. Astron. Soc.  {\bf 371}, 1373 (2006)
\bibitem{Wein03}
N. N., Weinberg, \&, M. Kamionkowski, Astrophys. J., {\bf 341}, 251, (2003)
\bibitem{Mota04}
D. F. Mota \&, C. van de Bruck C., Astronomy \& Astrophysics,
{\bf 421}, 71, (2004)

\bibitem{cham1}
  P.~Brax, C.~van de Bruck, A.~C.~Davis, J.~Khoury and A.~Weltman,
  Phys.\ Rev.\  D {\bf 70}, 123518 (2004)

\bibitem{cham2} D.~F.~Mota and J.~D.~Barrow,
  Mon.\ Not.\ Roy.\ Astron.\ Soc.\  {\bf 349}, 291 (2004)

\bibitem{cham3} P.~Brax, C.~van de Bruck, A.~C.~Davis and A.~M.~Green,
  Phys.\ Lett.\  B {\bf 633}, 441 (2006)

\bibitem{cham4} D.~F.~Mota and D.~J.~Shaw,
  Phys.\ Rev.\  D {\bf 75}, 063501 (2007)

\bibitem{Horel05}
C. Horellou \& J. Berge, Mon. Not. Roy. Astron.  Soc., {\bf 360}, 1393, (2005)
\bibitem{Zeng05}
Ding-fang Zeng, \&, Yi-hong Gao, 2005, (astro-ph/0505164)
\bibitem{Maor05}
I. Maor, \&, O. Lahav, Journal of Cosmology and
  Astroparticle Physics, {\bf 7}, 3, (2005)
\bibitem{david} David F. Mota JCAP {\bf 09}, 006 (2008)
\bibitem{Perc05}
W. J. Percival, Astronomy \& Astrophysics, {\bf 443}, 819, (2005)
\bibitem{Nunes06}
N. J. Nunes, \&, D. F. Mota, Mon. Not. Roy. Astron. Soc., {\bf 368}, 751, (2006)

\bibitem{Wang06}
P. Wang, Astrophys. J., {\bf 640}, 18, (2006)

\bibitem{Basi07}
S. Basilakos, \&, N. Voglis,
Mon. Not. Roy. Astron. Soc., {\bf 374}, 269, (2007)


%\cite{Broadhurst:2008re}
\bibitem{Broadhurst:2008re}
  T.~Broadhurst, K.~Umetsu, E.~Medezinski, M.~Oguri and Y.~Rephaeli,
  %``Comparison of Cluster Lensing Profiles with Lambda CDM Predictions,''
  Astrophys.\ J.\  {\bf 685}, L9 (2008)
  [arXiv:0805.2617 [astro-ph]].
  %%CITATION = ASJOA,685,L9;%%



\bibitem{Linjen03}
E. V. Linder and A. Jenkins, Mon. Not. Roy. Astron. Soc.,
{\bf 346}, 573, (2003)

\bibitem{Linder2004}
E.~V.~Linder, Phys.\ Rev.\ Lett.\  {\bf 70}, 023511, (2004);
E.~V.~Linder, Rep.\ Prog.\ Phys., {\bf 71}, 056901, (2008)

\bibitem{Eis05}
D. J. Eisenstein D. J., et al., Astrophys. J., {\bf 633}, 560, (2005);
N. Padmanabhan, et al., Mon. Not. Roy. Astron. Soc., {\bf 378}, 852, (2007)

\bibitem{Bond:1997wr}
  J.~R.~Bond, G.~Efstathiou and M.~Tegmark,
  Mon.\ Not.\ Roy. Astron. Soc.\  {\bf 291}, L33, (1997)

\bibitem{Trotta:2004qj}
  R.~Trotta,
  %``Cosmic Microwave Background Anisotropies: Beyond Standard Parameters,''
  [arXiv:astro-ph/0410115].
  %%CITATION = ASTRO-PH 0410115;%%

\bibitem{Nesseris:2006er}
  S.~Nesseris, \&,  L.~Perivolaropoulos,
  %``Crossing the phantom divide: Theoretical implications and observational
  %status,''
  JCAP {\bf 0701}, 018, (2007)

\bibitem{Deff}
C. Deffayet, G. Dvali, \&, G. Cabadadze, Phys. Rev. D., {\bf 65}, 044023, (2002)
\bibitem{Chevallier:2001qy}
M.~Chevallier and D.~Polarski,
Int.\ J.\ Mod.\ Phys.\ D {\bf 10}, 213, (2001)

\bibitem{Linder:2002et}
E.~V.~Linder,
Phys.\ Rev.\ Lett.\  {\bf 90}, 091301, (2003)

\bibitem{Linder2007}
E.~V.~Linder, \&, R. N. Cahn, Astrop. \ Phys., {\bf 28}, 481, (2007)

\bibitem{Cont2007}
I. Contopoulos, \&, S. Basilakos, Astron. Astrophys. {\bf 471}, 59, (2007)

\bibitem{Alam2003}
U. Alam, V. Sahni, T. D. Saini, A. A. Starobinsky,
Mon. Not. Roy. Astron. Soc., {\bf 344}, 1057, (2003)

\bibitem{Sorbo2007}
J. A. Frieman, C.T  Hill, A. Stebbins, I. Waga, Phys. Rev. Lett., {\bf 75}, 2077, (1995);
K. Dutta, \&, L. Sorbo, Phys. Rev. D., {\bf 75}, 063514, (2007);
A. Abrahamse, A. Albrecht, M. Bernard, B. Bozek, Phys. Rev. D., {\bf 77}, 103504, (2008)

\bibitem{Doran2006}
I. Zlatev, L. Wang, P. J. Steinhardt, Phys. Rev. Lett., {\bf 82}, 896, (1999);
M. Doran, S. Stern, E. Thommes, JCAP, {\bf 0704}, 015, (2006)

\bibitem{Bento03}
M. C. Bento, O. Bertolami, A. A. Sen, Phys. Rev. D., {\bf 70}, 083519, (2004)

\bibitem{Guo05}
Z. K. Guo, \&, Y. Z., Zhang, astro-ph/0506091, (2005);
Z. K. Guo, \&, Y. Z., Zhang, astro-ph/050979, (2005)

\bibitem{Vcgdata}
M. Makler, S. Q. de Oliveira, I. Waga, Phys. Lett B., {\bf 555}, 1, (2003);
M. C. Bento, O. Bertolami, A. A. Sen, Phys. Lett. B., {\bf 575}, 172, (2003);
A. Dev, J. S. Alcaniz, D. Jain, Phys. Rev. D., {\bf 67}, 023515, (2003);
Y. Gong, C. K. Duan, Mon. Not. Roy. Astron. Soc., {\bf 352}, 847, (2004);
Z. H. Zhu, Astron. Astrophys., {\bf 423}, 421, (2004)
L. Amendola, I. Waga, F. Finelli, astro-ph/0509099;

\bibitem{Kirk03}
D. Kirkman, D. Tytler, N. Suzuki, N., J. M. O'Meara, D. Lubin,
Astrophys. J. Suppl., ApJS, {\bf 149}, 1, (2003)

\bibitem {Peeb93}
P. J. E. Peebles, 1993, Principles of Physical Cosmology,
Princeton University Press, Princeton New Jersey, (1993)

%46
\bibitem {Stab06}
F. H. Stabenau, \& B. Jain, Phys. Rev. D, {\bf 74}, 084007, (2006)

%47
\bibitem {Uzan07}
P. J.  Uzan, Gen.\ Rel.\ Grav., {\bf 39}, 307, (2007)

\bibitem{Silv94}
V. Silveira, \& I. Waga, Phys. Rev. D., {\bf 64}, 4890, (1994)

%51
\bibitem{Nes08}
S. Nesseris \&  L. Perivolaropoulos, Phys. Rev D., {\bf 77}, 3504, (2008)

%52
\bibitem{Linca08}
V. E. Linder \& N. R. Cahn, Astroparticle Physics, {\bf 28}, 481,
(2007)

%53
\bibitem{Gann08}
R. Gannouji, \&, D. Polarski,  (2008), [arXiv:0802.4196]

\bibitem{Gunn72}
J. E. Gunn, \&, J. R. Gott, Astrophys. J., {\bf 176}, 1, (1972)

\bibitem{Mul05}
C. R. Mullis, P. Rosati, G. Lamer, H. B$\ddot {\rm o}$ehringer,
P. Schuecker \&, R. Fassbender,
Mon. Not. Roy. Astron. Soc., {\bf 623}, L85, (2005);
S. A. Stanford, et al., Astrophys. J., {\bf 646}, L13, (2006)

\bibitem{Kita96}
T. Kitayama, \&, Y. Suto, {\bf 469}, 480, (1996)

\bibitem{shaw} Douglas J. Shaw, David F. Mota, Astrophys.J.Suppl.,  {\bf 174}, 277, (2008).

\bibitem{Peeb84}
P. J. E. Peebles, Astrophys. J., {\bf 284}, 439, (1984);
S. Weinberg, {\bf 59}, 2607, (1987);

\bibitem{Rich92}
D. Richstone, A. Loeb \&, E. L. Turner, Astrophys. J., {\bf 393}, 477, (1992)


\bibitem{press}  W.~H.~Press and P.~Schechter,
  Astrophys.\ J.\  {\bf 187}, 425 (1974).

\bibitem{eke}
V. Eke, S. Cole \&, C. S. Frenk, Mon. Not. Roy. Astron. Soc., {\bf 282}, 263, (1996)

\bibitem{Shaef08}
B. M. Schafer, \&, K. Koyama, Mon. Not. Roy. Astron. Soc,  {\bf 385}, 411, (2008)

\bibitem{Fran08}
%M. J. Francis, G. F. Lewis, E. V. Linder, Mon. Not. Roy. Astron. Soc,  {\bf 393}, L31, (2009)
M. Bartelmann, M. Doran, C. Wetterich,  Astron. \& Astrophys. , {\bf 454}, 27, (2006)

%\cite{Navarro:1995iw}
\bibitem{Navarro:1995iw}
  J.~F.~Navarro, C.~S.~Frenk and S.~D.~M.~White,
  %``The Structure of Cold Dark Matter Halos,''
  Astrophys.\ J.\  {\bf 462}, 563 (1996)
  [arXiv:astro-ph/9508025].
  %%CITATION = ASJOA,462,563;%%

%\cite{Navarro:1996gj}
\bibitem{Navarro:1996gj}
  J.~F.~Navarro, C.~S.~Frenk and S.~D.~M.~White,
  %``A Universal Density Profile from Hierarchical Clustering,''
  Astrophys.\ J.\  {\bf 490}, 493 (1997)
  [arXiv:astro-ph/9611107]
  %%CITATION = ASJOA,490,493;%%

%\cite{Wechsler:2001cs}
\bibitem{Wechsler:2001cs}
  R.~H.~Wechsler, J.~S.~Bullock, J.~R.~Primack, A.~V.~Kravtsov and A.~Dekel,
  %``Concentrations of Dark Halos from their Assembly Histories,''
  Astrophys.\ J.\  {\bf 568}, 52 (2002)
  [arXiv:astro-ph/0108151].
  %%CITATION = ASJOA,568,52;%%

%\cite{Umetsu:2007pq}
\bibitem{Umetsu:2007pq}
  K.~Umetsu and T.~Broadhurst,
  %``Combining Lens Distortion and Depletion to Map the Mass Distribution of
  %A1689,''
  Astrophys.\ J.\  {\bf 684}, 177 (2008)
  [arXiv:0712.3441]
  %%CITATION = ASJOA,684,177;%%

%\cite{Miyazaki:2002wa}
\bibitem{Miyazaki:2002wa}
  S.~Miyazaki {\it et al.},
  %``Subaru Prime Focus Camera -- Suprime-Cam --,''
  arXiv:astro-ph/0211006.
  %%CITATION = ASTRO-PH/0211006;%%

%\cite{Eke:2000av}
\bibitem{Eke:2000av}
  V.~R.~Eke, J.~F.~Navarro and M.~Steinmetz,
  %``The Power Spectrum Dependence of Dark Matter Halo Concentrations,''
  Astrophys.\ J.\  {\bf 554}, 114 (2001)
  [arXiv:astro-ph/0012337].
  %%CITATION = ASJOA,554,114;%%


%\cite{Mota:2008ne}
\bibitem{Mota:2008ne}
  D.~F.~Mota,
  %``Probing Dark Energy at Galactic and Cluster Scales,''
  [arXiv:0812.4493 [astro-ph]].
  %%CITATION = JCAPA,09,;%%


%\cite{Liberato:2006un}
\bibitem{Liberato:2006un}
  L.~Liberato and R.~Rosenfeld,
  %``Dark energy parameterizations and their effect on dark halos,''
  JCAP {\bf 0607}, 009 (2006)
  [arXiv:astro-ph/0604071].
  %%CITATION = JCAPA,0607,009;%%

%\cite{Nesseris:2007pa}
\bibitem{Nesseris:2007pa}
  S.~Nesseris and L.~Perivolaropoulos,
  %``Testing LCDM with the Growth Function \delta(a): Current Constraints,''
  Phys.\ Rev.\  D {\bf 77}, 023504 (2008)
  [arXiv:0710.1092 [astro-ph]].
  %%CITATION = PHRVA,D77,023504;%%

%\cite{Dave:2002mn}
\bibitem{Dave:2002mn}
  R.~Dave, R.~R.~Caldwell and P.~J.~Steinhardt,
  %``Sensitivity of the cosmic microwave background anisotropy to initial
  %conditions in quintessence cosmology,''
  Phys.\ Rev.\  D {\bf 66} (2002) 023516
  [arXiv:astro-ph/0206372].


\end{thebibliography}
\end{document}